\documentclass[11pt,twoside]{article}
\usepackage{asp2010}

\resetcounters

\begin{document}

\title{A search for star-planet interactions in chromospheric lines}
\author{Lenz, L. F.$^1$, Reiners, A.$^1$, and K\"urster, M.$^2$
\affil{$^1$Universit\"at G\"ottingen, Institut f\"ur Astrophysik, Friedrich-Hund-Platz 1, 37077 Goettingen, Germany}
\affil{$^2$Max Planck Institut f\"ur Astronomie, K\"onigstuhl 17
69117 Heidelberg, Germamy}}

\begin{abstract}
 Massive planets in very close orbits around their central stars can induce  so-called star-planet interactions (SPI), which may be of magnetic or gravitational nature. In both cases, SPI can potentially cause recurring chromospheric emission on the host star visible in Ca~\textsc{ii}~H~\&~K and/or H$\alpha$. The emission would be bound to the planetary orbit, not to the rotation period of the star. 
We searched for SPI in a sample of 7 stars with massive close-in planets using high-resolution spectroscopic data taken at HRS (HET) and FEROS (La Silla). We find no periodically recurring emission in the planet-hosting stars. In the case of HD 41004 AB, a binary system consisting of a K dwarf and an M dwarf, where the M dwarf is orbited by a brown dwarf companion, we find signs of cyclic variation in the Ca~\textsc{ii}~K and H$\alpha$ emission lines that could be associated to interactions between the M dwarf and its companion. We present our first results of this interesting system that may become an important system for the understanding of SPI.
\end{abstract}

\section{Introduction}
Star-planet interactions (SPI) in stellar chromospheres could occur as recurring enhanced Ca \textsc{ii} H \& K flux following the periodicity of the planet orbiting its host star. So far signs for SPI were found only in few systems and the empirical evidence is still heavyly debated \citep{popp, shkol1, scharf}. 
The possible physical scenarios for SPI are gravitational and/or magnetic interactions. Gravitational interaction in the system could lead to tidal bulges on the surface of the star, changing the local geometry. It is perhaps conceivable that this may
in turn favor magnetic reconnection thereby enhancing the observed stellar
activity. Tidal interaction would lead to two enhancement peaks over the planetary phase. In the case of magnetic interaction, the reconnection of field lines from the planet and the star could lead to enhanced stellar activity peaking once per orbit \citep{cuntz}. 
Recent observations for HD 179949 even suggest an ``On/Off" behavior: after some years the periodicity of the Ca \textsc{ii} K emission changed from the planetary to the stellar rotation period \citep{shkol1}.

\section{Observations}
\begin{figure}[!ht]
\plotone{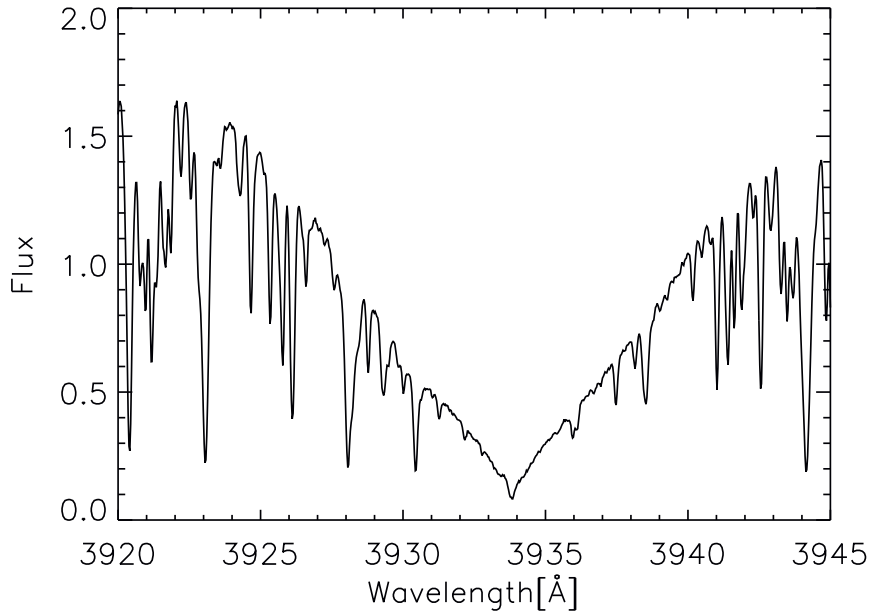}
\caption{Ca \textsc{ii} K line in a spectrum of HD 75289 observed with FEROS at the 2.2 ESO/MPG Telescope, La Silla}
\label{fig:spec}
\end{figure}

We obtained FEROS (La Silla) and HRS (HET) spectra for a sample of 7 stars, each of which has a verified massive close-in planet. The spectra have a S/N of ca. 30 spectral pixel$^{-1}$ (FEROS) and 90 spectral pixel$^{-1}$ (HRS) in the Ca \textsc{ii} K core. The parameters of the systems are listed in Table \ref{table}. A typical spectrum is shown in Fig.\ref{fig:spec}. 

\begin{table}[!ht]
\caption{Parameters of the program stars}
\label{table}
\smallskip
\begin{center}
{\small
\begin{tabular}{lcccccc}
\tableline
\noalign{\smallskip}
Star & Spectral-Type & $d$ & $M$ & $P\rm_{orb}$ & Spectrograph & \# \\
      &    &[AU]& [M$\rm_{Jup}$]& [d]		&	  & spectra \\
\noalign{\smallskip}
\tableline
\noalign{\smallskip}
$\upsilon$And 	& F8 V & 0.059 & 0.69 & 4.62 & HRS & 11\\
$\tau$Boo	& F7 V & 0.046 & 3.9 & 3.31 & HRS & 12 \\
HD 217101	& G8 IV & 0.073 & 1.33 & 7.12 & FEROS & 7 \\
HD 212301	& F8 V & 0.036 & 0.45 & 2.457 & FEROS & 6\\
HD 75289	& G0 V & 0.046 & 0.42 & 3.51 & FEROS & 7 \\
Gl 876		& M4 V & 0.021 & 0.02 & 1.94 & FEROS & 4 \\
HD 41004AB	& M2 & 0.018 & 18.4 & 1.33 & FEROS & 9 \\ 
\noalign{\smallskip}
\tableline
\end{tabular}
}
\end{center}
\end{table}

FEROS spectra were reduced with the FEROS reduction pipeline. A continuum normalization was applied to the reduced spectra, using two normalization regions in the pseudo continuum left and right of the Ca~\textsc{ii}~K line core.

The HRS data reduction was performed with the MIDAS software following standard reduction techniques for optical spectra except for the normalization with flatfields. The HRS flatfields cannot be used for the Ca~\textsc{ii}~K region since the separate calibration fibre is wider than the science fiber and the flatfield orders overlap in the region around 4000\AA. Thus we reduced the data without flatfielding. For our differential analysis this is not crucial. Normalization of the HRS spectra was performed by fixing the flux of each spectrum to unity at 3929.5\AA. To remove the underlying flux distribution each spectrum was divided by a reference spectrum which was the first observed spectrum of all spectra from one target. A polynomial was fitted to this quotient representing the missing blaze function. Finally each spectrum was divided by its fit for normalization.

\section{Results}
\begin{figure}[!ht]
\plottwo{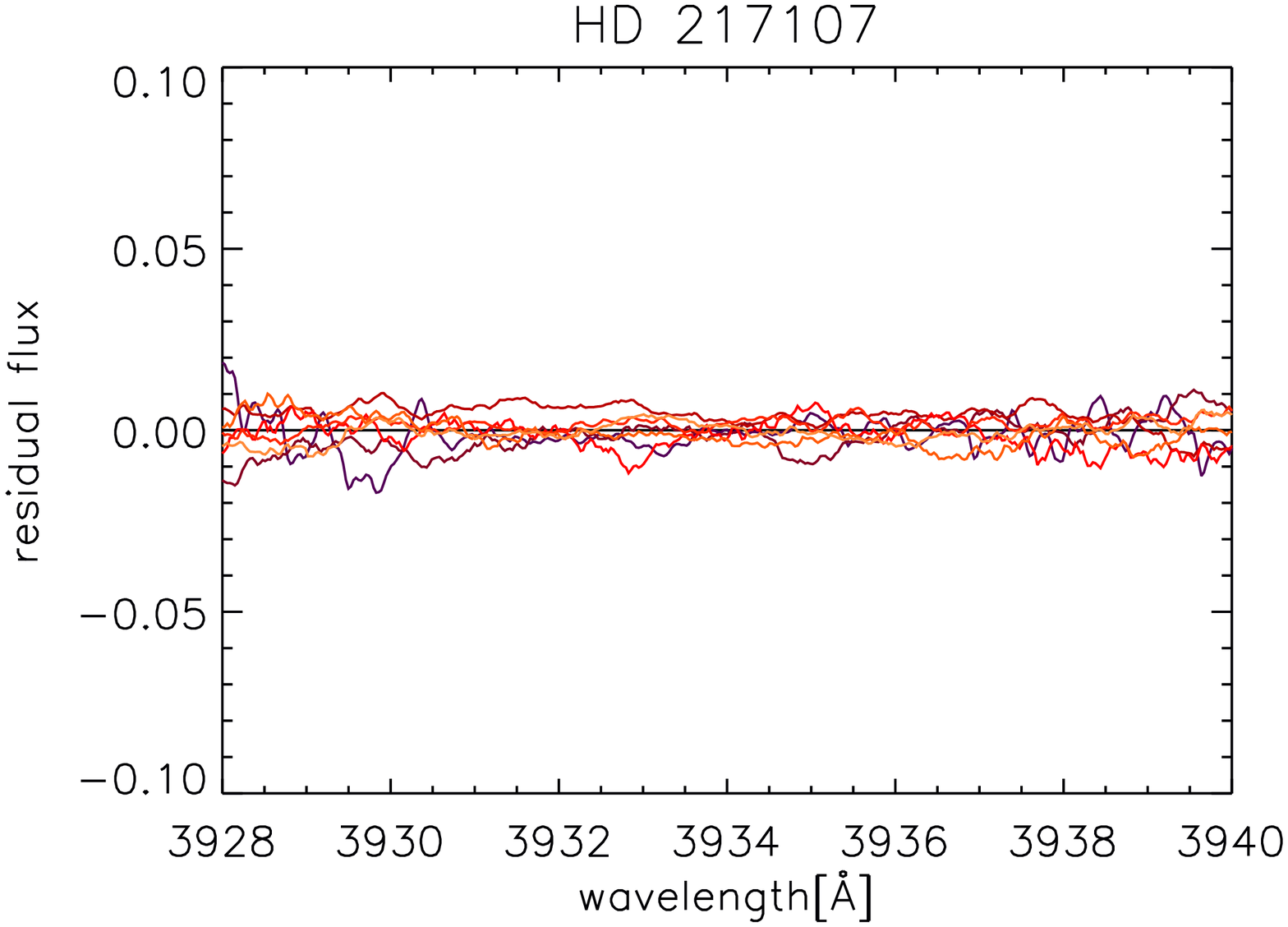}{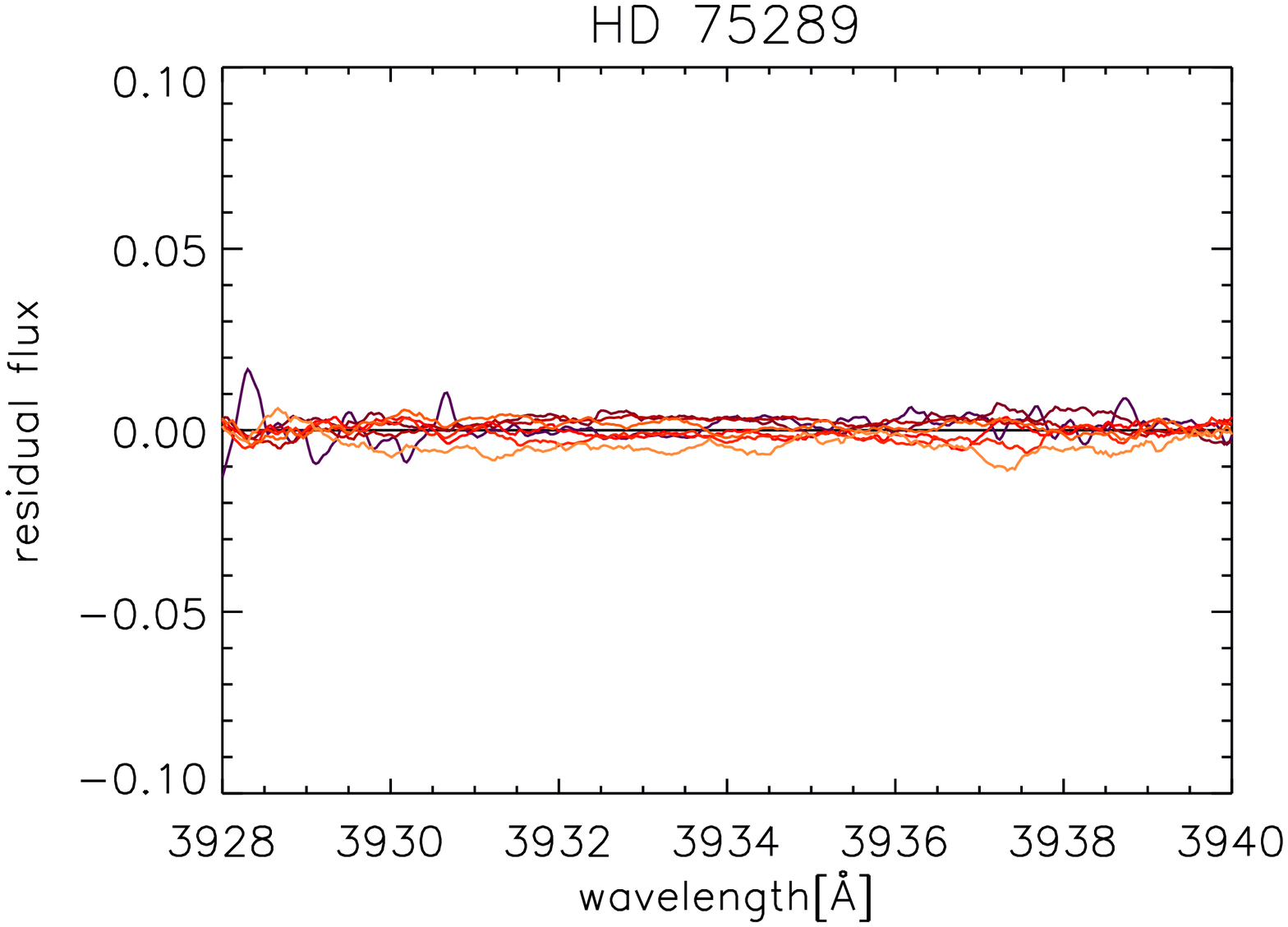}
\caption{Ca \textsc{ii} residuals for two of the program stars.}
\label{fig:residuals}
\end{figure}

To analyse the spectra for variations in the Ca~\textsc{ii}~K and H$\alpha$ lines we cross correlate all spectra of one target in order to correct for spectral shifts due to the barycentric Earth movement and spectrograph instabilities and compute the respective mean spectrum for each dataset. We subtract these mean spectra binwise from each spectrum to receive the residual fluxes. Variable emission in the line core due to chromospheric activity would result in a bulge in the residual spectrum. For most of our target stars we find no variability in the line residuals that could be associated with SPI. Planetary masses are between 0.02 and 3.9~M$\rm_{Jup}$, all planets orbit their hosts at distances closer than 0.08~AU, orbital periods are shorter than 5~d.
The data quality of both spectrographs should allow us to detect variations in the residuals on the order of those reported in \cite{shkol1}. We typically took 6 or more observations per star at a cadence of one spectrum per night for the FEROS data and one spectrum every 6 nights for the HET data. Fig.\ref{fig:residuals} shows two typical residual plots of the target stars HD~75289 and HD~217107.

\begin{figure}[!ht]
\plottwo{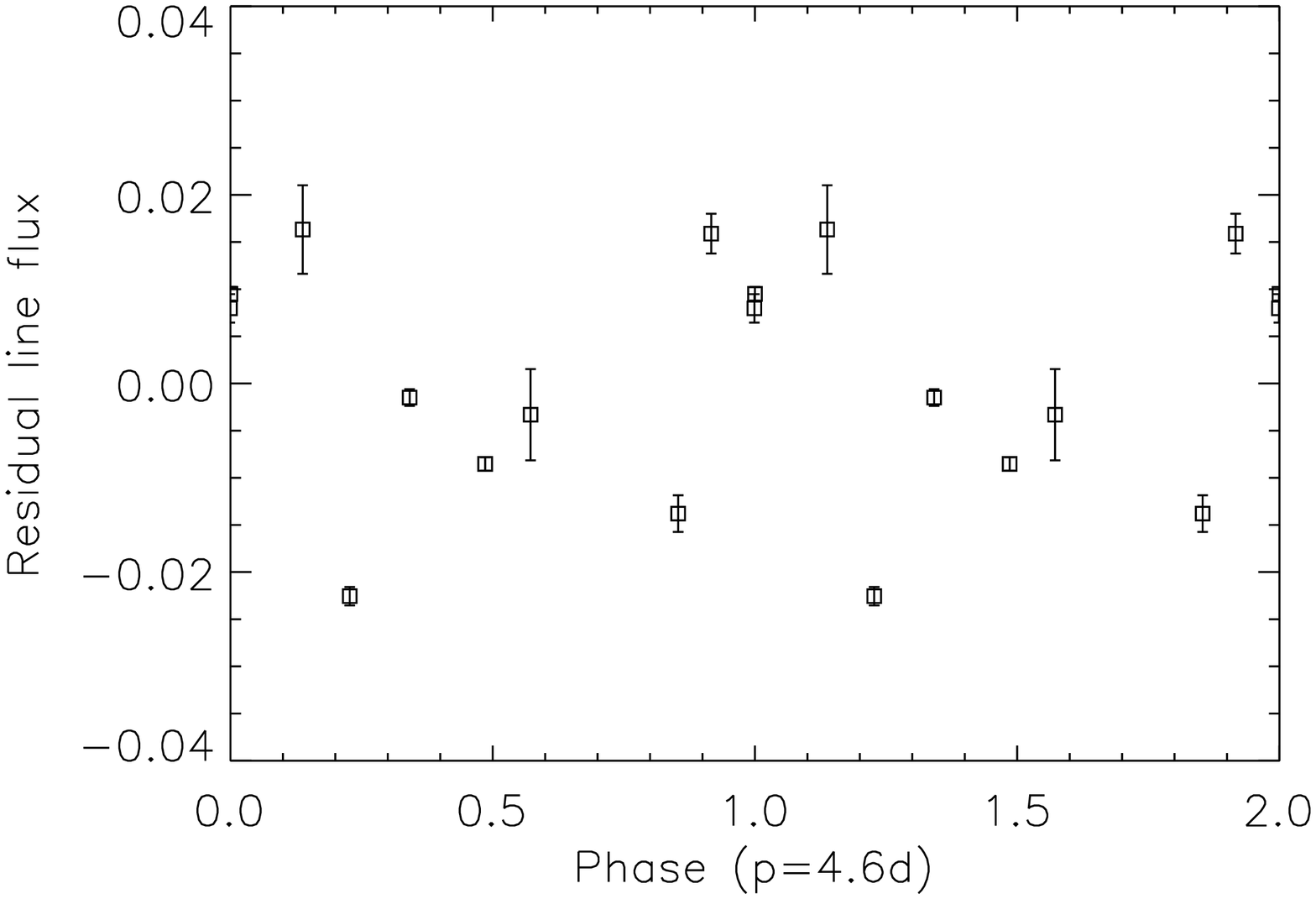}{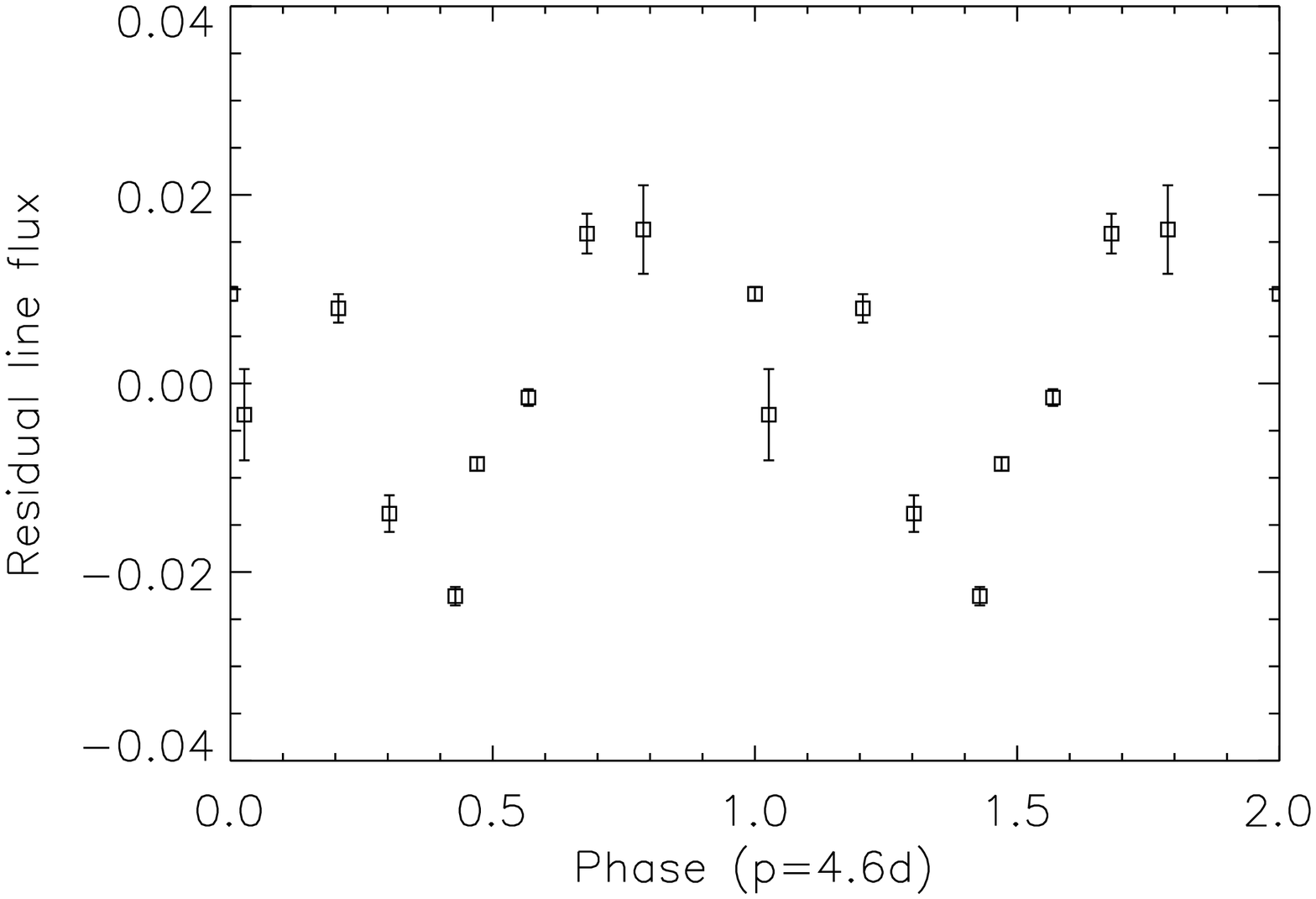}
\caption{$\upsilon$And: total residual Ca \textsc{ii} K line flux folded with the planetary rotation period (p$_{orb}$=4.6d) and the estimated rotation period of 9.5d}
\label{fig:upsints}
\end{figure}

The $\upsilon$And spectra observed with the HRS at HET show variations in the Ca~\textsc{ii}~K flux. The integrated flux residuals plotted over the orbital phase of the planet ($p\rm_{orb}$=4.6~d) give no significant variation that would indicate a dependence between the chromospheric activity and the orbiting planet. Combining our data with FOCES data of $\upsilon$ And \citep{popp2} indicates a period of the variability of 9.5 days which is close to the rotation period of the star. Fig.\ref{fig:upsints} gives an overview of the total residual flux plotted over both periods.

\subsection{A brown dwarf companion to HD 41004B}

\begin{figure}[!ht]
\plottwo{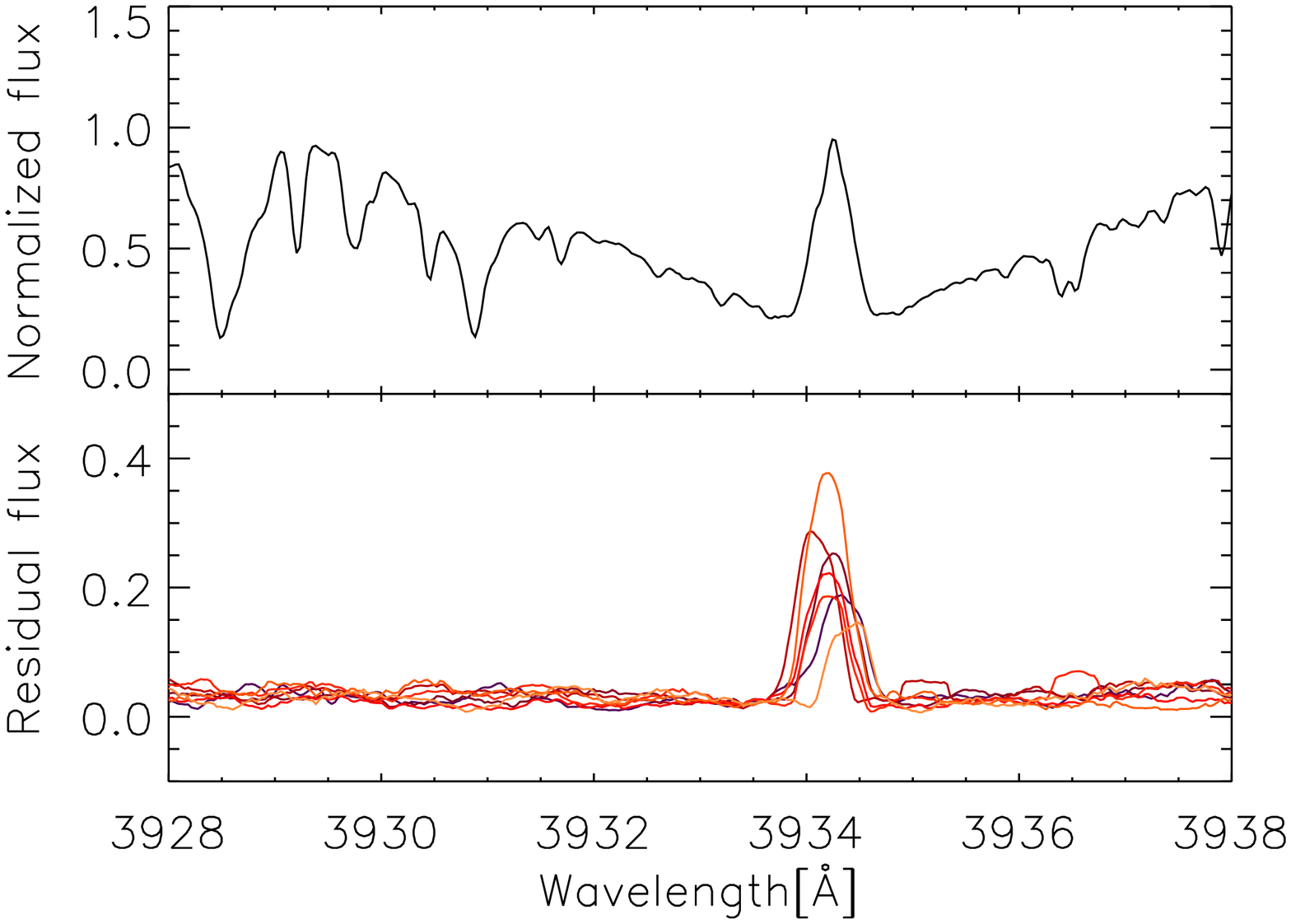}{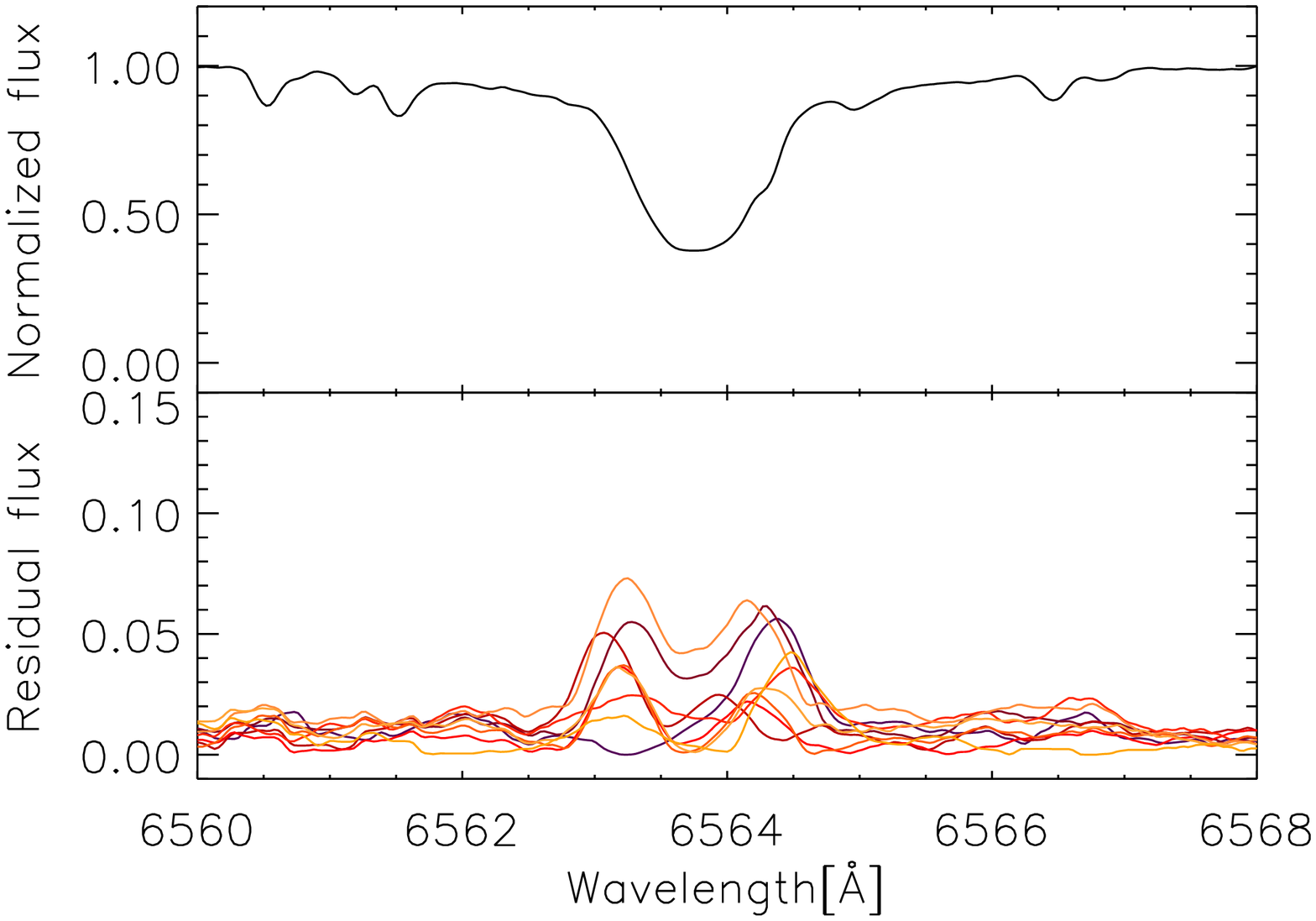}
\caption{left: Variability in Ca \textsc{ii} K line cores of the HD 41004 AB FEROS data. \textit{upper panel:} minimum spectrum computed from the normalized spectra; \textit{lower panel:} residual flux; right: variability in H$\alpha$ line cores of the HD 41004 AB FEROS data. \textit{upper panel:} minimum spectrum computed from the normalized spectra; \textit{lower panel:} residual flux }
\label{fig:camulti}
\end{figure}

HD 41004AB is a binary system of a K dwarf and a M dwarf with a separation of 0.5". We could not separate the two stars so that our spectra contain light of both components. Furthermore the B component (the M dwarf) is orbited by a brown dwarf companion ($M\rm_{BD}$=18.4~M$\rm_{Jup}$, $P\rm_{orb}$=1.33~d). The Ca~\textsc{ii}~K and H $\alpha$ lines show variation in the residual plots. For the residual plots for HD 41004AB we subtract a minimal spectrum calculated from all spectra instead of a mean spectrum. The residual plots are shown in Fig\ref{fig:camulti}.

\begin{figure}[!ht]
\plottwo{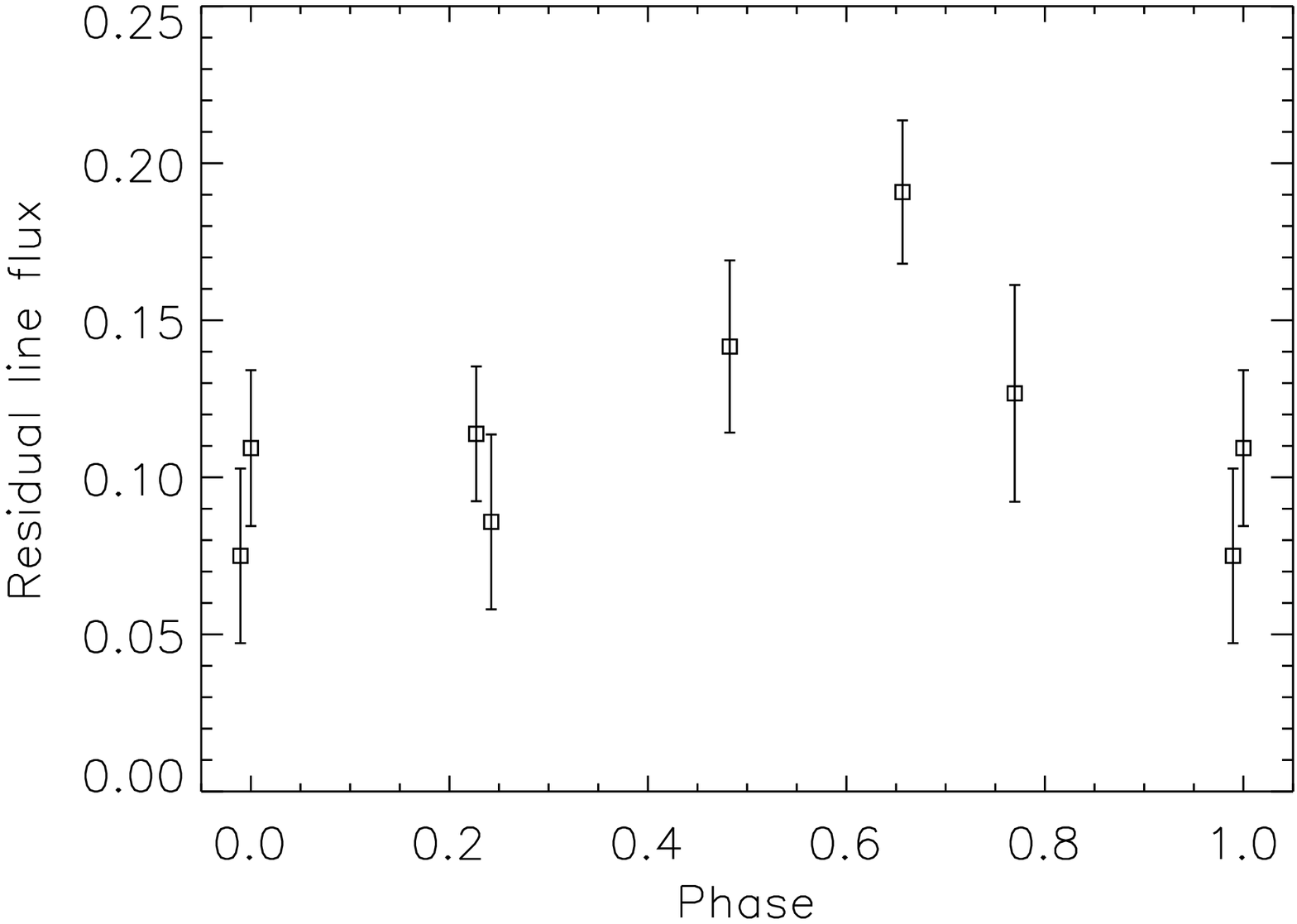}{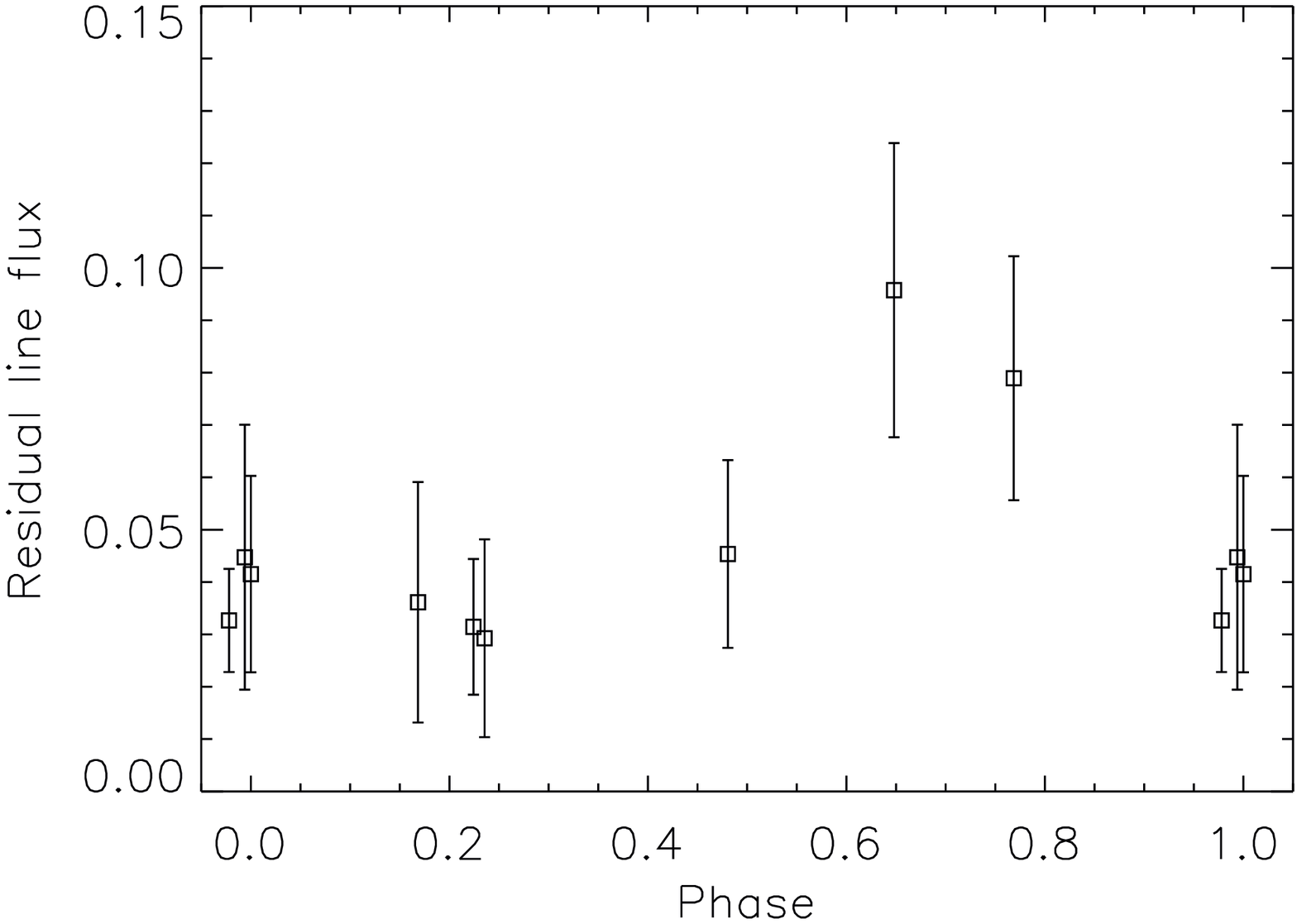}
\caption{HD 41004 AB: \textit{left:} Total residual Ca \textsc{ii} K line flux folded with the orbital period of the brown dwarf (p$_{orb}$=1.33d); \textit{right}: Total residual H$\alpha$ line flux folded with the same period}
\label{fig:hdays}
\end{figure}

We analyze nine spectra of this system, which where recorded in the nights between Nov 28 until Dec 10.  Two of the spectra for HD 41004 AB have cosmics close to the  Ca~\textsc{ii}~K line. Thus they are only used for the analysis of the H$\alpha$ line and excluded from the Ca~\textsc{ii}~K line analysis. The residuals show variability in both the Ca~\textsc{ii}~K and the H$\alpha$ lines.
At this point it is not clear whether the variability is caused by interactions of the M dwarf and its brown dwarf companion. Alternative scenarios include intrinsic variability of the A and/or B component and gravitational forces of the binary system. Nevertheless, the short timescale of the variation is consistent with the brown dwarf's orbital period and renders Interaction of the brown dwarf and its host star a very likely explanation. The integrated line flux of the Ca~\textsc{ii}~K and H$\alpha$ residuals folded with the orbital period of the brown dwarf is shown in Fig. \ref{fig:hdays}. So far, our data set is too small to present a meaningful period analysis.
However, our data are consistent with a variation of both lines with
the 1.33d period and in phase with each other. It is therefore quite
likely that we have found star-companion interactions in this system.

\section{Conclusions}
We find no flux variations in phase with the planetary orbit in the chromospheric Ca~\textsc{ii}~K and H$\alpha$ lines in the spectra of our sample stars that have close-in planets. The detected variability of $\upsilon$And is probably due to the rotation of the star and not bound to the planet.

HD 41004 AB potentially is a very good candidate to observe interactions since the mass-ratio of the M dwarf and its brown dwarf companion is between that of interacting binaries and star-planet systems. More spectra will be obtained in the future to search for periodicities of the flux variations.

\acknowledgements 
A.R. acknowledges research funding from the DFG under RE 1664/4-1.
The Hobby-Eberly Telescope (HET) is a joint project of the University of Texas at Austin, the Pennsylvania State University, Stanford University, Ludwig-Maximilians-Universit\"at M\"unchen, and Georg-August-Universit\"at G\"ottingen. The HET is named in honor of its principal benefactors, William P. Hobby and Robert E. Eberly.

\bibliography{lenz_l}

\begin{thebibliography}{}
\expandafter\ifx\csname natexlab\endcsname\relax\def\natexlab#1{#1}\fi
\expandafter\ifx\csname url\endcsname\relax
  \def\url#1{\texttt{#1}}\fi
\expandafter\ifx\csname urlprefix\endcsname\relax\def\urlprefix{URL }\fi
\providecommand{\eprint}[2][]{\url{#2}}

\bibitem[{Cuntz et~al.(2000)Cuntz, Saar, \& Musielak}]{cuntz}
Cuntz, M., Saar, S.~H., \& Musielak, Z.~E. 2000, ApJ, 533, 151

\bibitem[{Poppenhaeger et~al.(2010{\natexlab{a}})Poppenhaeger, Lenz, Reiners,
  \& Schmitt}]{popp2}
Poppenhaeger, K., Lenz, L.~F., Reiners, A., \& Schmitt, J.~H.~M.~M.
  2010{\natexlab{a}}, ArXiv e-prints. \eprint{1010.5632}

\bibitem[{Poppenhaeger et~al.(2010{\natexlab{b}})Poppenhaeger, Robrade, \&
  Schmitt}]{popp}
Poppenhaeger, K., Robrade, J., \& Schmitt, J.~H.~M.~M. 2010{\natexlab{b}},
  A\&A, 515, 98

\bibitem[{Scharf(2010)}]{scharf}
Scharf, C.~A. 2010, ApJ, 722, 1547

\bibitem[{Shkolnik et~al.(2008)Shkolnik, Bohlender, Walker, \&
  Cameron}]{shkol1}
Shkolnik, E., Bohlender, D.~A., Walker, G.~A.~H., \& Cameron, A.~C. 2008, ApJ,
  676, 628

\end{thebibliography}

\end{document}